\documentclass[final,5p,times]{elsarticle}

\usepackage{amsfonts}
\usepackage{eurosym}
\usepackage{amssymb,bbold}
\usepackage{amsmath}
\usepackage{natbib}
\usepackage{mathrsfs}
\usepackage{hyperref}

\hypersetup{
  colorlinks=true,linkcolor=blue,citecolor=blue,
  filecolor=blue,urlcolor=blue,breaklinks=true
}

\biboptions{sort&compress}
\journal{Physics Letters B}
\begin{document}

\begin{frontmatter}

\title{On the $\kappa$-Dirac Oscillator revisited}

\author[uepg]{F. M. Andrade}
\ead{fmandrade@uepg.br}
\author[ufma]{E. O. Silva}
\ead{edilbertoos@pq.cnpq.br}
\author[ufma]{M. M. Ferreira Jr.}
\ead{manojr.ufma@gmail.com}
\author[ufma]{E. C. Rodrigues}
\ead{ednilson.fisica@gmail.com}

\address[uepg]{
  Departamento de Matem\'{a}tica e Estat\'{i}stica,
  Universidade Estadual de Ponta Grossa,
  84030-900 Ponta Grossa-PR, Brazil
}
\address[ufma]{
  Departamento de F\'{i}sica,
  Universidade Federal do Maranh\~{a}o,
  Campus Universit\'{a}rio do Bacanga,
  65085-580 S\~{a}o Lu\'{i}s-MA, Brazil
}

\begin{abstract}
This Letter is based on the $\kappa$-Dirac equation, derived from the
$\kappa$-Poincar\'{e}-Hopf algebra.
It is shown that the $\kappa$-Dirac equation preserves parity while
breaks charge conjugation and time reversal symmetries.
Introducing the Dirac oscillator prescription,
$\mathbf{p}\to\mathbf{p}-im\omega\beta\mathbf{r}$, in the $\kappa$-Dirac
equation, one obtains the $\kappa$-Dirac oscillator.
Using a decomposition in terms of spin angular functions, one achieves
the deformed radial equations, with the associated deformed energy
eigenvalues and eigenfunctions.
The deformation parameter breaks the infinite degeneracy of the Dirac
oscillator.
In the case where $\varepsilon=0$, one recovers the energy eigenvalues
and eigenfunctions of the Dirac oscillator.
\end{abstract}


\end{frontmatter}

\section{Introduction}
\label{sec:introduction}

In 1989, it was proposed in a seminal paper by Moshinsky and Szczepaniak
\cite{JPA.1989.22.817} the basic idea of a relativistic quantum
mechanical oscillator, called Dirac oscillator.
Such oscillator behaves as an harmonic oscillator with a strong
spin-orbit coupling in the non-relativistic limit.
Since the time of its proposal it has been the object of considerable
attention in various branches of theoretical physics.
For instance, it appears in mathematical physics
\cite{PLA.2004.325.21,PLB.2007.644.311,JPA.1997.30.2585,
JPA.2005.38.1747,JPA.2007.40.6427,JPA.1991.24.667,
PLB.2012.710.478,AP.2013.336.489,JPA.2010.43.285204,PLA.2003.311.93},
nuclear physics
\cite{PLA.2012.376.3475,PRC.2012.85.054617,AP.2005.320.71},
quantum optics
\cite{JOB.2002.4.R1,OL.2010.35.1302,EPJB.2012.85.237,PRA.2007.76.041801},
supersymmetry \cite{JPA.1995.28.6447,JPA.2006.39.10909,CTP.2008.49.319},
and noncommutativity
\cite{PLA.2012.376.2467,IJTP.2013.52.441,IJTP.2012.51.2143,IJMPA.2011.26.4991}.
Recently, the first experimental realization of the Dirac oscillator was
realized by J. A. Franco-Villafa\~{n}e \textit{et al.}
\cite{PRL.2013.111.170405}, which should draw even more attention for
such system.
Moreover, C. Quibay \textit{et al.} proposed that the Dirac oscillator
can describe some electronic properties of monolayer and bylayer
graphene \cite{arXiv:1311.2021} and show the existence of a 
quantum phase transition in this system \cite{arXiv:1312.5251}.

The Dirac oscillator has also been discussed in connection with the
theory of quantum deformations \cite{EPL.1997.39.583}.
Some of these deformations are based on the $\kappa$-deformed
Poincar\'{e}-Hopf algebra, with $\kappa$ being a masslike fundamental
deformation parameter, introduced in
Refs. \cite{PLB.1991.264.331,PLB.1992.293.344} and further discussed in
Refs. \cite{PLB.1993.302.419,PLB.1993.318.613,PLB.1994.329.189,PLB.1994.334.348}.
The $\kappa$-deformed algebra is defined by the following commutation
relations:
\begin{subequations}
  \label{eq:algebra}
  \begin{align}
    \left[ p_{\nu },p_{\mu }\right]  = {} & 0,  \label{eq:algebraa} \\
    \left[ M_{i},p_{\mu }\right]  = {} &(1-\delta_{0\mu })i\epsilon_{ijk}p_{k}, \\
    \left[ L_{i},p_{\mu }\right]  = {} &
    i[p_{i}]^{\delta_{0\mu }}[\delta_{ij}\varepsilon^{-1}
    \sinh \left(\varepsilon p_{0}\right) ]^{1-\delta_{0\mu }}, \\
    \left[ M_{i},M_{j}\right]  = {} & i\epsilon_{ijk}M_{k},\qquad
    \left[ M_{i},L_{j} \right] =i\epsilon_{ijk}L_{k}, \\
    \left[ L_{i},L_{j}\right]  = {} & -i\epsilon_{ijk}\left[ M_{k}\cosh \left(
        \varepsilon p_{0}\right) -\frac{\varepsilon ^{2}}{4}p_{k}p_{l}M_{l}\right] ,
  \end{align}
\end{subequations}
where $\varepsilon $ is defined by
\begin{equation}
\varepsilon=\kappa^{-1}=\lim_{R\rightarrow \infty }(R\ln q),
\end{equation}
with $R$ being the de Sitter curvature, $q$ is a real deformation
parameter, and $p_{\mu}=(p_{0},\mathbf{p})$ is the $\kappa$-deformed
generator for energy and momenta.
Also, the $M_{i}$, $L_{i}$ represent the spatial rotations and deformed
boosts generators, respectively.
The coalgebra and antipode for the $\kappa$-deformed Poincar\'{e}-Hopf
algebra was established in Ref. \cite{AoP.1995.243.90}.

Several investigations have been developed in the latest years in the
context of this theoretical framework on space-like $\kappa$-deformed
Minkowski spacetime.
The interest in this issue also appears in field theories
\cite{CQG.2010.27.025012,PLB.2012.711.122,NPB.2001.102-103.161,EPJC.2003.31.129},
quantum electrodynamics
\cite{PLB.2002.529.256,PRD.2011.84.085020,JHEP.2011.1112.080},
realizations in terms of commutative coordinates and derivatives
\cite{EPJC.2013.73.2472,PRD.2009.79.045012,EPJC.2006.47.531,EPJC.2008.53.295},
relativistic quantum systems
\cite{PLB.2013.719.467,CQG.2004.21.2179,JHEP.2004.2004.28,PLB.1995.359.339,
PLB.1994.339.87}, doubly special relativity
\cite{PRD.2013.87.125009}, noncommutative black holes
\cite{PRD.2012.85.045029} and the construction of scalar theory
\cite{PRD.2009.80.025014}.

The aim of this letter is to suitably describe the $\kappa$-Dirac
oscillator making use of the $\kappa$-Poincar\'{e}-Hopf algebra, tracing
a comparison with the results of Ref. \cite{EPL.1997.39.583}, where it
was argued that usual approach for introducing the Dirac oscillator,
$\mathbf{p}\to\mathbf{p}-im\omega \beta  \mathbf{r}$, in the
$\kappa$-Dirac equation \cite{PLB.1993.302.419,PLB.1993.318.613},
has not led to the Dirac oscillator spectrum in the limit $\varepsilon
\to 0$.
This result, however, contradicts the well-known fact that the
$\kappa$-Dirac equation recovers the standard Dirac equation in this
limit.
In this context, this letter reassessed the $\kappa$-Dirac oscillator
problem yielding a modified oscillator spectrum that indeed regains the
Dirac oscillator behavior in the limit $\varepsilon\to 0$.

The plan of our Letter is the following.
In Section \ref{sec:k-dirac} we introduce the $\kappa$-Dirac
analyzing its behavior under $\mathcal{C}$, $\mathcal{P}$, $\mathcal{T}$
(discrete) symmetries.
In Section  \ref{sec:k-dirac-oscillator} the oscillator prescription is
implemented in order to study the physical implications of the
$\kappa$-deformation in the Dirac oscillator problem.
Using a decomposition in terms of spin angular functions, we write the
relevant radial equation to study the dynamics of the system.
The Section \ref{sec:eigen} is devoted to the calculation the
energy eigenvalues and eigenfunctions of the $\kappa$-Dirac oscillator
and to the discussion of the results.
A brief conclusion in outlined in Section \ref{sec:conclusion}.

\section{$\kappa$-Dirac equation and discrete symmetries}
\label{sec:k-dirac}

In this section, we present  $\kappa$-Dirac equation, invariant under
the $\kappa$-Poincar\'{e} quantum algebra \cite{PLB.1993.302.419},
considering $O(\varepsilon)$ \cite{PLB.1993.318.613}:
\begin{equation}\label{eq:deformed-Dirac}
  \left\{
    (\gamma_{0}p_{0}-\gamma_{i}p_{i}) +
    \frac{\varepsilon}{2}
    \left[
      \gamma_{0}\left( p_{0}^{2}-p_{i}p_{i}\right) -mp_{0}
    \right]
  \right\} \psi = m\psi.
\end{equation}
which recovers the standard Dirac equation in the limit
$\varepsilon\to 0$.

An initial discussion refers to the behavior of this deformed equation
under $\mathcal{C}$, $\mathcal{P}$, $\mathcal{T}$ (discrete) symmetries.
Concerning the parity operator ($\mathcal{P}$), in the context of the
Dirac equation, $\mathcal{P}=i\gamma^{0}$, with
$\mathcal{P}\gamma^{\mu}\mathcal{P}^{-1}=\gamma_{\mu}$ and
$\psi_{P}=\mathcal{P}\psi$ being the parity-transformed spinor.
Applying $\mathcal{P}$ on the Dirac deformed equation, we attain
\begin{equation}
  \left\{
    (\gamma_{0}p_{0}-\gamma_{i}p_{i})+\frac{\varepsilon}{2}
    \left[\gamma_{0}\left(p_{0}^{2}-p_{i}p_{i}\right)-mp_{0}
    \right]
  \right\}
  \psi_{\mathcal{P}}= m\psi_{\mathcal{P}},
\end{equation}
concluding that it is invariant under $\mathcal{P}$  action.

We can now verify that this equation is not invariant under charge
conjugation ($\mathcal{C}$) and time reversal ($\mathcal{T}$).
As for the $\mathcal{C}$ operation, the charge-conjugated spinor is
$\psi_{\mathcal{C}}=U_{\mathcal{C}}\psi^{*}=\mathcal{C}\gamma^{0}\psi^{*}$,
with $\mathcal{C}=i\gamma^{2}\gamma^{0}$ being the charge conjugation
operator, and $U_{\mathcal{C}}\gamma^{\mu\ast}U_{\mathcal{C}}^{-1}=-\gamma^{\mu}$.
On the other hand, the time reversal operator is,
$\mathcal{T}=i\gamma^{1}\gamma^{3}$, so that
$\psi_{\mathcal{T}}(x,t^{\prime})=\mathcal{T}\psi^{*}(x,t^{\prime})$,
and $\mathcal{T}\gamma^{\mu\ast}\mathcal{T}^{-1}=(\gamma^{0},-\gamma^{i})$.
Applying $U_{\mathcal{C}}$ and $\mathcal{T}$ on the complex conjugate of
Eq.\eqref{eq:deformed-Dirac}, we achieve:
\begin{align}
  \left\{
    (\gamma_{0}p_{0}-\gamma_{i}p_{i})+\frac{\varepsilon}{2}
    \left[-\gamma_{0}\left( p_{0}^{2}-p_{i}p_{i}\right)-mp_{0}
    \right]
  \right\}
  \psi_{\mathcal{C}}   = {} & m\psi_{\mathcal{C}},\\
  \left\{
    (\gamma_{0}p_{0}-\gamma_{i}p_{i})+\frac{\varepsilon}{2}
    \left[
      \left(\gamma_{0}\right)\left(p_{0}^{2}-p_{i}p_{i}\right)+mp_{0}
    \right]
  \right\}  \psi_{\mathcal{T}}   = {} & m\psi_{\mathcal{T}}.
\end{align}
Theses equations differ from Eq. \eqref{eq:deformed-Dirac},
revealing that the $\mathcal{C}$ and $\mathcal{T}$ are not symmetries
of this system.
As a consequence, particle and anti-particle eigenenergies should
become different.
Further, note that under $\mathcal{CT}$ or $\mathcal{CPT}$ operations
the original equation is modified as
\begin{equation}
  \left\{
    (\gamma_{0}p_{0}-\gamma_{i}p_{i})-\frac{\varepsilon}{2}
    \left[
      \gamma_{0}\left(p_{0}^{2}-p_{i}p_{i}\right)-mp_{0}
    \right]
  \right\}
  \psi^{\prime}= m\psi^{\prime},
\end{equation}
where $\psi^{\prime}=\psi_{\mathcal{CT}}$ or
$\psi^{\prime}=\psi_{\mathcal{CPT}}$, showing that this equation is not
invariant under $\mathcal{CT}$ or $\mathcal{CPT}$ operations, once the
parameter $\varepsilon$ is always positive.

\section{$\kappa$-Dirac oscillator equation}
\label{sec:k-dirac-oscillator}

Now, we derive the equation that governs the dynamics of the
Dirac oscillator in the context of Eq. \eqref{eq:deformed-Dirac}.
The Dirac oscillator stems from the prescription \cite{JPA.1989.22.817}
\begin{subequations}\label{eq:coupling}
  \begin{align}
    p_{0}\rightarrow {}& p_{0}=H_{0}, \label{eq:couplinga} \\
    \mathbf{p}\rightarrow {}& \mathbf{p}-i m \omega \beta \mathbf{r},
    \label{eq:couplingb}
  \end{align}
\end{subequations}
where $\mathbf{r}$ is the position vector, $m$ is the mass of particle
and $\omega$ the frequency of the oscillator.
The $\kappa$-Dirac oscillator can be obtained by substituting
Eq. \eqref{eq:coupling} into Eq. \eqref{eq:deformed-Dirac}.
The result is
\begin{equation}\label{eq:eigen}
  H\psi =E\psi,
\end{equation}
with
\begin{equation}\label{eq:Hkdirac}
  H=H_{0}-
  \frac{\varepsilon}{2}
  \left[
    p_{0}^{2}-
    (\mathbf{p}-i m \omega \beta \mathbf{r})
    (\mathbf{p}-i m \omega \beta \mathbf{r})
    -\beta mp_{0}\right],
\end{equation}
where $H_{0}$ represents the undeformed part of the Dirac operator
\begin{equation}\label{eq:hundeformed}
  H_{0}=\boldsymbol{\alpha}\cdot 
  \left( \mathbf{p}-i m \omega \beta \mathbf{r}\right) +\beta m.  
\end{equation}
At this point it is important trace a comparison with the
results of Ref. \cite{EPL.1997.39.583}, in which it is argued that the
prescription of the Eq. \eqref{eq:coupling}, yielding the
$\kappa$-deformed Hamiltonian of Eq. \eqref{eq:Hkdirac}, does not lead
to an oscillator-like spectrum even when  $\varepsilon\to 0$.
This result, however, is not correct, as properly shown in
Section \ref{sec:eigen}.
Furthermore, another deformed wave equation is introduced without any
kind of proof (see Eq. (15) in \cite{EPL.1997.39.583}).
Here, instead of postulating a deformed wave equation, we follow a
pragmatic approach obtaining the $\kappa$-Dirac oscillator equation
\eqref{eq:Hkdirac} from basic principles.

In the four-dimensional representation, the matrices $\boldsymbol{\gamma}$
and $\boldsymbol{\alpha}$ are given by
\begin{equation}
  \beta =\left(
    \begin{array}{cc}
      I & 0 \\
      0 & -I%
    \end{array}
  \right),
  ~~~\boldsymbol{\gamma}=\beta \boldsymbol{\alpha}\mathbf{=}\left(
    \begin{array}{cc}
      0 & \boldsymbol{\sigma} \\
      -\boldsymbol{\sigma} & 0
    \end{array}%
  \right) ,~~\ \boldsymbol{\alpha =}\left(
    \begin{array}{cc}
      0 & \boldsymbol{\sigma} \\
      \boldsymbol{\sigma} & 0
    \end{array}%
  \right) ,~~  \label{eq:mgamma}
\end{equation}
and obey the anticommutation relations and the square identity,
\begin{equation*}
  \left\{ \alpha_{i},\alpha_{j}\right\} =0,~~~i\neq j
\end{equation*}%
\begin{equation*}
  \left\{ \alpha_{i},\beta \right\} =0,
\end{equation*}
\begin{equation*}
  \alpha_{i}^{2}=\beta ^{2}=I.
\end{equation*}
In the representation \eqref{eq:mgamma}, $\psi $ may be written as a
bispinor $\psi =\left(\psi_{1},\psi_{2}\right)^{T}$ in terms of
two-component spinors $\psi_{1}$ and $\psi_{2}$.
Thus, Eq. \eqref{eq:eigen} leads to
\begin{multline}\label{eq:edoa}
  \left(1+\frac{m\varepsilon}{2}\right)
  \left(\boldsymbol{\sigma}\cdot \boldsymbol{\pi}^{+}\right) \psi_{2}
  =
  \left(E-m\right) \psi_{1} \\
  +\varepsilon
  \left[
    im\omega \left( \mathbf{r}\cdot \mathbf{p}\right)
    +
    m\omega \left( \boldsymbol{\sigma}\cdot \mathbf{L}\right)
    +
    m^{2}\omega^{2}r^{2}
  \right] \psi_{1},
\end{multline}
\begin{multline}\label{eq:edob}
  \left(1-\frac{m\varepsilon}{2}\right)
  \left(\boldsymbol{\sigma}\cdot \boldsymbol{\pi}^{-}\right) \psi_{1}
  =
  \left(E+m\right) \psi_{2}   \\
  -\varepsilon
  \left[
    im\omega \left( \mathbf{r}\cdot \mathbf{p}\right)
    +
    m\omega \left( \boldsymbol{\sigma}\cdot \mathbf{L}\right)
    -
    m^{2}\omega^{2}r^{2}
  \right] \psi_{2},
\end{multline}
where
\begin{equation}  \label{eq:pi}
\boldsymbol{\pi}^{\pm}=\mathbf{p}\pm i m \omega \mathbf{r}.
\end{equation}
Since we are interested in studying the $\kappa$-Dirac oscillator in a
three-dimensional spacetime, Eqs. \eqref{eq:edoa} and \eqref{eq:edob} above
may be solved in spherical coordinates.
First, using the property
\begin{equation}
  \boldsymbol{\sigma}\cdot\mathbf{p}=
  \left(\boldsymbol{\sigma}\cdot \mathbf{\hat{r}}\right)
  \left(
    \mathbf{\hat{r}}\cdot \mathbf{p}+
    i\frac{\boldsymbol{\sigma}
      \cdot \mathbf{L}}{r}
  \right),
\end{equation}
with
$\boldsymbol{\sigma}\cdot\mathbf{r}=
r\boldsymbol{\sigma}\cdot\mathbf{\hat{r}}$,
we rewrite the quantity $\boldsymbol{\sigma}\cdot\boldsymbol{\pi}^{\pm}$
as
\begin{equation}  \label{eq:idd}
  \boldsymbol{\sigma}\cdot\boldsymbol{\pi}^{\pm}=
  (\boldsymbol{\sigma}\cdot \boldsymbol{\hat{r}})
  \left(
    \mathbf{\hat{r}}\cdot \mathbf{p}+
    i\frac{\mathbf{\hat{K}}-\mathbf{1}}{r}\pm i m \omega r
  \right),
\end{equation}
where the operator $\mathbf{\hat{K}}$ is related to the orbital angular
momentum operator $\mathbf{\hat{L}}$ as
\begin{equation}\label{eq:opk}
  \mathbf{\hat{K}}=\boldsymbol{\sigma}\cdot \mathbf{\hat{L}}+\mathbf{1}.
\end{equation}
We seek solutions of the form
\begin{equation}
  \psi =
  \left(
    \begin{array}{c}
      \psi_{1}(\mathbf{x}) \\
      \psi_{2}(\mathbf{x})
    \end{array}
  \right) =
  \left(
    \begin{array}{c}
       f(r)\chi_{k}^{m_{j}}(\theta,\phi)  \\
      ig(r)\chi_{-k}^{m_{j}}(\theta ,\phi )
    \end{array}
  \right) = \frac{1}{r}
  \left(
    \begin{array}{c}
       u(r)\chi_{k}^{m_{j}}(\theta ,\phi)  \\
      iv(r)\chi_{-k}^{m_{j}}(\theta ,\phi)
    \end{array}
  \right) ,  \label{eq:wavef}
\end{equation}
where $\chi_{\pm k}^{m_{j}}(\theta ,\phi)$ are the spin angular
functions \cite{Book.1998.Strange}, with
\begin{equation}
  k=
  \begin{cases}
    -(\ell +1) , & \mbox{for}j=\ell +1/2, \\
    \ell, & \mbox{for}j=\ell -1/2.
  \end{cases}
\end{equation}
By substituting Eq. \eqref{eq:wavef} into Eqs.
\eqref{eq:edoa} and \eqref{eq:edob}, and using the relations
\begin{align}
  \left(\boldsymbol{\sigma}\cdot\mathbf{\hat{r}}\right)\chi_{\pm k}^{m_{j}}
   = {} & -\chi_{\mp k}^{m_{j}}, \\
  \mathbf{\hat{K}}\chi_{\pm k}^{m_{j}}
  = {} & \mp k\chi_{\pm k}^{m_{j}},
\end{align}
we find a set of two coupled radial differential equations of first
order:
\begin{multline}
  \left(1+\frac{m\varepsilon}{2}\right)
  \left[
    -v^{\prime}+\left( \frac{k}{r}+m\omega r\right) v
  \right] =
  \varepsilon m\omega ru^{\prime}   \\
  +\left\{\left( E-m\right) +
    \varepsilon \left[ m\omega k+m^{2}\omega^{2}r^{2}\right] \right\} u,
\end{multline}
\begin{multline}
  \left(1-\frac{m\varepsilon}{2}\right)
  \left[
    u^{\prime}+\left( \frac{k}{r}+m\omega r\right) u
  \right]
  =-\varepsilon m\omega  rv^{\prime}  \\
  +\left\{
    \left(E+m\right) +\varepsilon
    \left[
      m\omega k+m^{2}\omega^{2}r^{2}
    \right]
  \right\} v.
\end{multline}
After some algebra, the above equations are decoupled yielding a single
second order equation for $u(r)$,
\begin{multline} \label{eq:edof}
  u^{\prime \prime}+2m^{2}\varepsilon \omega ru^{\prime}-
  \left[
    \frac{\ell\left( \ell +1\right)}{r^{2}}+
    \left(1-2 m \varepsilon\right) m^{2}\omega^{2}r^{2}-\mu_{\varepsilon}
  \right] u =0.
\end{multline}
A similar equation exists for $v(r)$.
Here
\begin{equation}\label{eq:mue}
  \mu_{\varepsilon}=
  (E^{2}-m^{2})-[(2k-1) (1+m\varepsilon )+\varepsilon E] m\omega,
\end{equation}
and we have used the result $k^{2}+k=\ell \left( \ell +1\right) $.

\section{Eigensolutions for the problem}
\label{sec:eigen}

In this section, we calculate the energy eigenvalues and eigenfunctions
of the $\kappa$-Dirac oscillator, making some comparisons with those in
the literature and discussing the associate results.
The regular solution for Eq. \eqref{eq:edof} is
\begin{align}\label{eq:sol1}
  u(r)  = {} &
  e^{-{m^2 \omega r^2}/{2}}
  \left[(1-m\varepsilon)m\omega r^2\right]^{{(\ell+1)}/{2}} \nonumber \\
  {} & \times
  M\left(\frac{1}{2}(\ell +\frac{3}{2}-a_{\varepsilon}),\ell+\frac{3}{2},
    (1-m\varepsilon) m \omega r^2\right),
\end{align}
with
\begin{equation}
  a_{\varepsilon}=
  \frac
  {\mu_{\varepsilon}-m^2\varepsilon \omega}
  {2(1-m\varepsilon)m\omega},
\end{equation}
and $M(a,b,z)$ is the confluent hypergeometric function of first kind
\cite{Book.1972.Abramowitz}.
The energy eigenvalues of the $\kappa$-Dirac oscillator come from
requiring that the first parameter in the confluent hypergeometric
function of Eq. \eqref{eq:sol1} is a negative integer,$-n$, with $n$ a
nonnegative integer.
By using $N=2n+\ell$ as principal quantum number, and with
$\mu_{\varepsilon}$ given by Eq. \eqref{eq:mue}, one finds
\begin{equation}\label{eq:energya}
  E^{2}-m^{2}=2m\omega
  \left[
    N+k+1+ m\varepsilon
    \left( \frac{E}{2m}-N+k-\frac{3}{2}\right)
  \right].
\end{equation}
By solving Eq. \eqref{eq:energya} for $E$, we obtain
\begin{align}\label{eq:energy}
  E_{\pm}  = {} & \pm
  \Big[
    \sqrt{
      2m\omega(N+k+1) +m^{2} +
      [2(k -N)-3] m^{2}\varepsilon\omega} \nonumber \\
    {} & \pm\frac{m\varepsilon}{2}\omega
    \Big],
\end{align}
which for $j=\ell+1/2$ implies
\begin{align}\label{eq:enery2a}
  E_{\pm} = {} & \pm
  \Big[
    \sqrt{
      2m\omega(N-j+1/2) +m^{2} -
      [2(j+N)+4] m^{2}\varepsilon\omega} \nonumber \\
    {} & \pm\frac{m\varepsilon}{2}\omega
    \Big],
\end{align}
and
\begin{align}\label{eq:enery2b}
  E_{\pm} = {} & \pm
  \Big[
  \sqrt{
    2m\omega(N+j+3/2) +m^{2} +
    [2(j-N)-2] m^{2}\varepsilon\omega} \nonumber \\
  {} &\pm\frac{m\varepsilon}{2}\omega
  \Big],
\end{align}
for $j=\ell-1/2$.
The fact that particle and anti-particle energies turn out to be
distinct, $E_{+}\neq E_{-},$ is a consequence of charge conjugation
symmetry breaking.

The limit $\varepsilon \to 0$ exactly conducts to the undeformed Dirac
oscillator \cite{Book.1998.Strange}, whose eigenenergies are
\begin{subequations}
\begin{align}
E_{\pm}  = {} &\pm\sqrt{2m\omega(N-j+1/2)+m^{2}},\\
E_{\pm}  = {} &\pm\sqrt{2m\omega(N+j+3/2)+m^{2}},
\end{align}
\end{subequations}
for $j=\ell+1/2$ and $j=\ell-1/2$, respectively.
These undeformed energy expressions yield an infinity degeneracy, once
for $j=l+1/2$ all states with $N\pm q$, $j\pm q$ have the same energy,
while for $j=l-1/2$ the equal energy states are the one with $N\pm q$,
$j\mp q$, being $q$ an integer.
This infinity degeneracy is now lifted by the terms involving the
deformation parameter, $\varepsilon$, inside the square root of Eqs.
\eqref{eq:enery2a} and \eqref{eq:enery2b}.
Note that, in the limit $\varepsilon \to 0$, the eigenfunction
\eqref{eq:sol1} also regains the undeformed Dirac oscillator counterpart
exhibited in \cite{Book.1998.Strange}, revealing the consistency of the
description here developed.

\section{Conclusions}
\label{sec:conclusion}

We have studied the $\kappa$-Dirac oscillator problem based on
the $\kappa$-deformed Poincar\'{e}-Hopf algebra and the
$\kappa$-Dirac equation.
First, we have analyzed the behavior of the $\kappa$-Dirac equation under
discrete symmetries.
Further, we have shown that the usual prescription
$\mathbf{p}\to\mathbf{p}-im\omega\beta\mathbf{r}$ leads to a modified
spectrum that in fact recovers the undeformed Dirac oscillator result.
Using a decomposition in terms of spin angular functions, we have
derived the deformed radial equation whose solution
has led to the deformed eigenenergies and eigenfunctions.
We have verified that the deformation parameter implies the breakdown of
charge conjugation, time reversal and $\mathcal{CPT}$ symmetries, while
preserving parity.
The deformation parameter modifies the energy eigenvalues and eigenfunctions
of the Dirac oscillator, breaking the infinite degeneracy of the energy
eigenvalues as well.

\section{Acknowledgments}

The authors would like to thank M. Pereira for helpful discussions.
E. O. Silva acknowledges research grant by CNPq-(Universal) project
No. 482015/2013-6 and FAPEMA-(Universal) project No. 00845/13.
F. M. Andrade acknowledges research grant by Fun\-da\-\c{c}\~{a}o
Arauc\'{a}ria project No. 205/2013, and M. M. Ferreira Jr is grateful to
FAPEMA, CAPES and CNPq for finantial support.

\bibliographystyle{model1a-num-names}

\begin{thebibliography}{57}
\expandafter\ifx\csname natexlab\endcsname\relax\def\natexlab#1{#1}\fi
\providecommand{\url}[1]{\texttt{#1}}
\providecommand{\href}[2]{#2}
\providecommand{\path}[1]{#1}
\providecommand{\DOIprefix}{doi:}
\providecommand{\ArXivprefix}{arXiv:}
\providecommand{\URLprefix}{URL: }
\providecommand{\Pubmedprefix}{pmid:}
\providecommand{\doi}[1]{\href{http://dx.doi.org/#1}{\path{#1}}}
\providecommand{\Pubmed}[1]{\href{pmid:#1}{\path{#1}}}
\providecommand{\bibinfo}[2]{#2}
\ifx\xfnm\relax \def\xfnm[#1]{\unskip,\space#1}\fi
\bibitem[{Moshinsky and Szczepaniak(1989)}]{JPA.1989.22.817}
\bibinfo{author}{M.~Moshinsky}, \bibinfo{author}{A.~Szczepaniak},
  \bibinfo{journal}{J. Phys. A} \bibinfo{volume}{22} (\bibinfo{year}{1989})
  \bibinfo{pages}{L817}. \DOIprefix\doi{10.1088/0305-4470/22/17/002}.
\bibitem[{Ferkous and Bounames(2004)}]{PLA.2004.325.21}
\bibinfo{author}{N.~Ferkous}, \bibinfo{author}{A.~Bounames},
  \bibinfo{journal}{Phys. Lett. A} \bibinfo{volume}{325} (\bibinfo{year}{2004})
  \bibinfo{pages}{21}. \DOIprefix\doi{10.1016/j.physleta.2004.03.033}.
\bibitem[{Kulikov et~al.(2007)Kulikov, Tutik, and
  Yaroshenko}]{PLB.2007.644.311}
\bibinfo{author}{D.~Kulikov}, \bibinfo{author}{R.~Tutik},
  \bibinfo{author}{A.~Yaroshenko}, \bibinfo{journal}{Phys. Lett. B}
  \bibinfo{volume}{644} (\bibinfo{year}{2007}) \bibinfo{pages}{311}.
  \DOIprefix\doi{10.1016/j.physletb.2006.11.068}.
\bibitem[{Toyama et~al.(1997)Toyama, Nogami, and Coutinho}]{JPA.1997.30.2585}
\bibinfo{author}{F.~M. Toyama}, \bibinfo{author}{Y.~Nogami},
  \bibinfo{author}{F.~A.~B. Coutinho}, \bibinfo{journal}{J. Phys. A}
  \bibinfo{volume}{30} (\bibinfo{year}{1997}) \bibinfo{pages}{2585}.
  \DOIprefix\doi{10.1088/0305-4470/30/7/034}.
\bibitem[{Quesne and Tkachuk(2005)}]{JPA.2005.38.1747}
\bibinfo{author}{C.~Quesne}, \bibinfo{author}{V.~M. Tkachuk},
  \bibinfo{journal}{J. Phys. A} \bibinfo{volume}{38} (\bibinfo{year}{2005})
  \bibinfo{pages}{1747}. \DOIprefix\doi{10.1088/0305-4470/38/8/011}.
\bibitem[{Akcay(2007)}]{JPA.2007.40.6427}
\bibinfo{author}{H.~Akcay}, \bibinfo{journal}{J. Phys. A} \bibinfo{volume}{40}
  (\bibinfo{year}{2007}) \bibinfo{pages}{6427}.
  \DOIprefix\doi{10.1088/1751-8113/40/24/010}.
\bibitem[{de~Lange(1991)}]{JPA.1991.24.667}
\bibinfo{author}{O.~L. de~Lange}, \bibinfo{journal}{J. Phys. A}
  \bibinfo{volume}{24} (\bibinfo{year}{1991}) \bibinfo{pages}{667}.
  \DOIprefix\doi{10.1088/0305-4470/24/3/025}.
\bibitem[{Pedram(2012)}]{PLB.2012.710.478}
\bibinfo{author}{P.~Pedram}, \bibinfo{journal}{Phys. Lett. B}
  \bibinfo{volume}{710} (\bibinfo{year}{2012}) \bibinfo{pages}{478}.
  \DOIprefix\doi{10.1016/j.physletb.2012.03.015}.
\bibitem[{Bakke and Furtado(2013)}]{AP.2013.336.489}
\bibinfo{author}{K.~Bakke}, \bibinfo{author}{C.~Furtado},
  \bibinfo{journal}{Ann. Phys. (NY)} \bibinfo{volume}{336}
  (\bibinfo{year}{2013}) \bibinfo{pages}{489}.
  \DOIprefix\doi{10.1016/j.aop.2013.06.007}.
\bibitem[{Sadurn\'{i} et~al.(2010)Sadurn\'{i}, Torres, and
  Seligman}]{JPA.2010.43.285204}
\bibinfo{author}{E.~Sadurn\'{i}}, \bibinfo{author}{J.~M. Torres},
  \bibinfo{author}{T.~H. Seligman}, \bibinfo{journal}{J. Phys. A}
  \bibinfo{volume}{43} (\bibinfo{year}{2010}) \bibinfo{pages}{285204}.
  \DOIprefix\doi{10.1088/1751-8113/43/28/285204}.
\bibitem[{Pacheco et~al.(2003)Pacheco, Landim, and Almeida}]{PLA.2003.311.93}
\bibinfo{author}{M.~Pacheco}, \bibinfo{author}{R.~Landim},
  \bibinfo{author}{C.~Almeida}, \bibinfo{journal}{Phys. Lett. A}
  \bibinfo{volume}{311} (\bibinfo{year}{2003}) \bibinfo{pages}{93--96}.
  \DOIprefix\doi{10.1016/S0375-9601(03)00467-5}.
\bibitem[{Mun\'{a}rriz et~al.(2012)Mun\'{a}rriz, Dom\'{i}nguez-Adame, and
  Lima}]{PLA.2012.376.3475}
\bibinfo{author}{J.~Mun\'{a}rriz}, \bibinfo{author}{F.~Dom\'{i}nguez-Adame},
  \bibinfo{author}{R.~Lima}, \bibinfo{journal}{Phys. Lett. A}
  \bibinfo{volume}{376} (\bibinfo{year}{2012}) \bibinfo{pages}{3475}.
  \DOIprefix\doi{10.1016/j.physleta.2012.10.029}.
\bibitem[{Grineviciute and Halderson(2012)}]{PRC.2012.85.054617}
\bibinfo{author}{J.~Grineviciute}, \bibinfo{author}{D.~Halderson},
  \bibinfo{journal}{Phys. Rev. C} \bibinfo{volume}{85} (\bibinfo{year}{2012})
  \bibinfo{pages}{054617}. \DOIprefix\doi{10.1103/PhysRevC.85.054617}.
\bibitem[{Faessler et~al.(2005)Faessler, Kukulin, and
  Shikhalev}]{AP.2005.320.71}
\bibinfo{author}{A.~Faessler}, \bibinfo{author}{V.~Kukulin},
  \bibinfo{author}{M.~Shikhalev}, \bibinfo{journal}{Ann. Phys. (NY)}
  \bibinfo{volume}{320} (\bibinfo{year}{2005}) \bibinfo{pages}{71}.
  \DOIprefix\doi{10.1016/j.aop.2005.05.008}.
\bibitem[{Dodonov(2002)}]{JOB.2002.4.R1}
\bibinfo{author}{V.~V. Dodonov}, \bibinfo{journal}{J. Opt. B: Quantum
  Semiclass. Opt.} \bibinfo{volume}{4} (\bibinfo{year}{2002})
  \bibinfo{pages}{R1}. \DOIprefix\doi{10.1088/1464-4266/4/1/201}.
\bibitem[{Longhi(2010)}]{OL.2010.35.1302}
\bibinfo{author}{S.~Longhi}, \bibinfo{journal}{Opt. Lett.} \bibinfo{volume}{35}
  (\bibinfo{year}{2010}) \bibinfo{pages}{1302}.
  \DOIprefix\doi{10.1364/OL.35.001302}.
\bibitem[{Wang et~al.(2012)Wang, Cao, and Xiong}]{EPJB.2012.85.237}
\bibinfo{author}{Y.~Wang}, \bibinfo{author}{J.~Cao},
  \bibinfo{author}{S.~Xiong}, \bibinfo{journal}{Eur. Phys. J. B}
  \bibinfo{volume}{85} (\bibinfo{year}{2012}) \bibinfo{pages}{237}.
  \DOIprefix\doi{10.1140/epjb/e2012-30243-7}.
\bibitem[{Bermudez et~al.(2007)Bermudez, Martin-Delgado, and
  Solano}]{PRA.2007.76.041801}
\bibinfo{author}{A.~Bermudez}, \bibinfo{author}{M.~A. Martin-Delgado},
  \bibinfo{author}{E.~Solano}, \bibinfo{journal}{Phys. Rev. A}
  \bibinfo{volume}{76} (\bibinfo{year}{2007}) \bibinfo{pages}{041801--}.
  \DOIprefix\doi{10.1103/PhysRevA.76.041801}.
\bibitem[{Moshinsky et~al.(1995)Moshinsky, Quesne, and
  Smirnov}]{JPA.1995.28.6447}
\bibinfo{author}{M.~Moshinsky}, \bibinfo{author}{C.~Quesne},
  \bibinfo{author}{Y.~F. Smirnov}, \bibinfo{journal}{J. Phys. A}
  \bibinfo{volume}{28} (\bibinfo{year}{1995}) \bibinfo{pages}{6447}.
  \DOIprefix\doi{10.1088/0305-4470/28/22/020}.
\bibitem[{Quesne and Tkachuk(2006)}]{JPA.2006.39.10909}
\bibinfo{author}{C.~Quesne}, \bibinfo{author}{V.~M. Tkachuk},
  \bibinfo{journal}{J. Phys. A} \bibinfo{volume}{39} (\bibinfo{year}{2006})
  \bibinfo{pages}{10909}. \DOIprefix\doi{10.1088/0305-4470/39/34/021}.
\bibitem[{Guo-Xing and Zhong-Zhou(2008)}]{CTP.2008.49.319}
\bibinfo{author}{J.~Guo-Xing}, \bibinfo{author}{R.~Zhong-Zhou},
  \bibinfo{journal}{Commun. Theor. Phys.} \bibinfo{volume}{49}
  (\bibinfo{year}{2008}) \bibinfo{pages}{319}.
\bibitem[{Mandal and Rai(2012)}]{PLA.2012.376.2467}
\bibinfo{author}{B.~P. Mandal}, \bibinfo{author}{S.~K. Rai},
  \bibinfo{journal}{Phys. Lett. A} \bibinfo{volume}{376} (\bibinfo{year}{2012})
  \bibinfo{pages}{2467}. \DOIprefix\doi{10.1016/j.physleta.2012.07.001}.
\bibitem[{Melo et~al.(2013)Melo, Montigny, Pompeia, and
  Santos}]{IJTP.2013.52.441}
\bibinfo{author}{G.~Melo}, \bibinfo{author}{M.~Montigny},
  \bibinfo{author}{P.~Pompeia}, \bibinfo{author}{E.~Santos},
  \bibinfo{journal}{Int. J. Theor. Phys.} \bibinfo{volume}{52}
  (\bibinfo{year}{2013}) \bibinfo{pages}{441}.
  \DOIprefix\doi{10.1007/s10773-012-1350-0}.
\bibitem[{Luo et~al.(2012)Luo, Wang, Li, and Jing}]{IJTP.2012.51.2143}
\bibinfo{author}{Z.-Y. Luo}, \bibinfo{author}{Q.~Wang},
  \bibinfo{author}{X.~Li}, \bibinfo{author}{J.~Jing}, \bibinfo{journal}{Int. J.
  Theor. Phys.} \bibinfo{volume}{51} (\bibinfo{year}{2012})
  \bibinfo{pages}{2143}. \DOIprefix\doi{10.1007/s10773-012-1094-x}.
\bibitem[{Maluf(2011)}]{IJMPA.2011.26.4991}
\bibinfo{author}{R.~V. Maluf}, \bibinfo{journal}{Int. J. Mod. Phys. A}
  \bibinfo{volume}{26} (\bibinfo{year}{2011}) \bibinfo{pages}{4991--5003}.
  \DOIprefix\doi{10.1142/S0217751X11054887}.
\bibitem[{Franco-Villafa\~{n}e et~al.(2013)Franco-Villafa\~{n}e, Sadurn\'{i},
  Barkhofen, Kuhl, Mortessagne, and Seligman}]{PRL.2013.111.170405}
\bibinfo{author}{J.~A. Franco-Villafa\~{n}e}, \bibinfo{author}{E.~Sadurn\'{i}},
  \bibinfo{author}{S.~Barkhofen}, \bibinfo{author}{U.~Kuhl},
  \bibinfo{author}{F.~Mortessagne}, \bibinfo{author}{T.~H. Seligman},
  \bibinfo{journal}{Phys. Rev. Lett.} \bibinfo{volume}{111}
  (\bibinfo{year}{2013}) \bibinfo{pages}{170405}.
  \DOIprefix\doi{10.1103/PhysRevLett.111.170405}.
\bibitem[{Quimbay and Strange(2013{\natexlab{a}})}]{arXiv:1311.2021}
\bibinfo{author}{C.~Quimbay}, \bibinfo{author}{P.~Strange}
  (\bibinfo{year}{2013}{\natexlab{a}}).
  \href{http://arxiv.org/abs/1311.2021}{\tt arXiv:1311.2021}.
\bibitem[{Quimbay and Strange(2013{\natexlab{b}})}]{arXiv:1312.5251}
\bibinfo{author}{C.~Quimbay}, \bibinfo{author}{P.~Strange}
  (\bibinfo{year}{2013}{\natexlab{b}}).
  \href{http://arxiv.org/abs/1312.5251}{\tt arXiv:1312.5251}.
\bibitem[{Ndimubandi(1997)}]{EPL.1997.39.583}
\bibinfo{author}{J.~Ndimubandi}, \bibinfo{journal}{Europhys. Lett.}
  \bibinfo{volume}{39} (\bibinfo{year}{1997}) \bibinfo{pages}{583}.
  \DOIprefix\doi{10.1209/epl/i1997-00398-1}.
\bibitem[{Lukierski et~al.(1991)Lukierski, Ruegg, Nowicki, and
  Tolstoy}]{PLB.1991.264.331}
\bibinfo{author}{J.~Lukierski}, \bibinfo{author}{H.~Ruegg},
  \bibinfo{author}{A.~Nowicki}, \bibinfo{author}{V.~N. Tolstoy},
  \bibinfo{journal}{Phys. Lett. B} \bibinfo{volume}{264} (\bibinfo{year}{1991})
  \bibinfo{pages}{331}. \DOIprefix\doi{10.1016/0370-2693(91)90358-W}.
\bibitem[{Lukierski et~al.(1992)Lukierski, Nowicki, and
  Ruegg}]{PLB.1992.293.344}
\bibinfo{author}{J.~Lukierski}, \bibinfo{author}{A.~Nowicki},
  \bibinfo{author}{H.~Ruegg}, \bibinfo{journal}{Phys. Lett. B}
  \bibinfo{volume}{293} (\bibinfo{year}{1992}) \bibinfo{pages}{344}.
  \DOIprefix\doi{10.1016/0370-2693(92)90894-A}.
\bibitem[{Nowicki et~al.(1993)Nowicki, Sorace, and Tarlini}]{PLB.1993.302.419}
\bibinfo{author}{A.~Nowicki}, \bibinfo{author}{E.~Sorace},
  \bibinfo{author}{M.~Tarlini}, \bibinfo{journal}{Phys. Lett. B}
  \bibinfo{volume}{302} (\bibinfo{year}{1993}) \bibinfo{pages}{419}.
  \DOIprefix\doi{10.1016/0370-2693(93)90419-I}.
\bibitem[{Biedenharn et~al.(1993)Biedenharn, Mueller, and
  Tarlini}]{PLB.1993.318.613}
\bibinfo{author}{L.~Biedenharn}, \bibinfo{author}{B.~Mueller},
  \bibinfo{author}{M.~Tarlini}, \bibinfo{journal}{Phys. Lett. B}
  \bibinfo{volume}{318} (\bibinfo{year}{1993}) \bibinfo{pages}{613}.
  \DOIprefix\doi{10.1016/0370-2693(93)90462-Q}.
\bibitem[{Lukierski and Ruegg(1994)}]{PLB.1994.329.189}
\bibinfo{author}{J.~Lukierski}, \bibinfo{author}{H.~Ruegg},
  \bibinfo{journal}{Phys. Lett. B} \bibinfo{volume}{329} (\bibinfo{year}{1994})
  \bibinfo{pages}{189}. \DOIprefix\doi{10.1016/0370-2693(94)90759-5}.
\bibitem[{Majid and Ruegg(1994)}]{PLB.1994.334.348}
\bibinfo{author}{S.~Majid}, \bibinfo{author}{H.~Ruegg}, \bibinfo{journal}{Phys.
  Lett. B} \bibinfo{volume}{334} (\bibinfo{year}{1994}) \bibinfo{pages}{348}.
  \DOIprefix\doi{10.1016/0370-2693(94)90699-8}.
\bibitem[{Lukierski et~al.(1995)Lukierski, Ruegg, and
  Zakrzewski}]{AoP.1995.243.90}
\bibinfo{author}{J.~Lukierski}, \bibinfo{author}{H.~Ruegg},
  \bibinfo{author}{W.~Zakrzewski}, \bibinfo{journal}{Ann. Phys. (NY)}
  \bibinfo{volume}{243} (\bibinfo{year}{1995}) \bibinfo{pages}{90}.
  \DOIprefix\doi{10.1006/aphy.1995.1092}.
\bibitem[{Arzano et~al.(2010)Arzano, Kowalski-Glikman, and
  Walkus}]{CQG.2010.27.025012}
\bibinfo{author}{M.~Arzano}, \bibinfo{author}{J.~Kowalski-Glikman},
  \bibinfo{author}{A.~Walkus}, \bibinfo{journal}{Class. Quantum Grav.}
  \bibinfo{volume}{27} (\bibinfo{year}{2010}) \bibinfo{pages}{025012}.
\bibitem[{Kovacevi\'{c} et~al.(2012)Kovacevi\'{c}, Meljanac, Pacho\l, and
  Strajn}]{PLB.2012.711.122}
\bibinfo{author}{D.~Kovacevi\'{c}}, \bibinfo{author}{S.~Meljanac},
  \bibinfo{author}{A.~Pacho\l}, \bibinfo{author}{R.~Strajn},
  \bibinfo{journal}{Phys. Lett. B} \bibinfo{volume}{711} (\bibinfo{year}{2012})
  \bibinfo{pages}{122}. \DOIprefix\doi{10.1016/j.physletb.2012.03.062}.
\bibitem[{Kosi\'{n}ski et~al.(2001)Kosi\'{n}ski, Lukierski, and
  Ma\'{s}lanka}]{NPB.2001.102-103.161}
\bibinfo{author}{P.~Kosi\'{n}ski}, \bibinfo{author}{J.~Lukierski},
  \bibinfo{author}{P.~Ma\'{s}lanka}, \bibinfo{journal}{Nucl. Phys. B - Proc.
  suppl.} \bibinfo{volume}{102} (\bibinfo{year}{2001}) \bibinfo{pages}{161}.
  \DOIprefix\doi{10.1016/S0920-5632(01)01552-3}.
\bibitem[{DimitrijeviÄ‡ et~al.(2003)DimitrijeviÄ‡, Jonke, MÃ¶ller,
  Tsouchnika, Wess, and Wohlgenannt}]{EPJC.2003.31.129}
\bibinfo{author}{M.~DimitrijeviÄ‡}, \bibinfo{author}{L.~Jonke},
  \bibinfo{author}{L.~MÃ¶ller}, \bibinfo{author}{E.~Tsouchnika},
  \bibinfo{author}{J.~Wess}, \bibinfo{author}{M.~Wohlgenannt},
  \bibinfo{journal}{Eur. Phys. J. C} \bibinfo{volume}{31}
  (\bibinfo{year}{2003}) \bibinfo{pages}{129}.
  \DOIprefix\doi{10.1140/epjc/s2003-01309-y}.
\bibitem[{Cougo-Pinto et~al.(2002)Cougo-Pinto, Farina, and
  Mendes}]{PLB.2002.529.256}
\bibinfo{author}{M.~Cougo-Pinto}, \bibinfo{author}{C.~Farina},
  \bibinfo{author}{J.~Mendes}, \bibinfo{journal}{Phys. Lett. B}
  \bibinfo{volume}{529} (\bibinfo{year}{2002}) \bibinfo{pages}{256}.
  \DOIprefix\doi{10.1016/S0370-2693(02)01253-4}.
\bibitem[{Harikumar et~al.(2011)Harikumar, Juri\'{c}, and
  Meljanac}]{PRD.2011.84.085020}
\bibinfo{author}{E.~Harikumar}, \bibinfo{author}{T.~Juri\'{c}},
  \bibinfo{author}{S.~Meljanac}, \bibinfo{journal}{Phys. Rev. D}
  \bibinfo{volume}{84} (\bibinfo{year}{2011}) \bibinfo{pages}{085020}.
  \DOIprefix\doi{10.1103/PhysRevD.84.085020}.
\bibitem[{DimitrijeviÄ‡ and Jonke(2011)}]{JHEP.2011.1112.080}
\bibinfo{author}{M.~DimitrijeviÄ‡}, \bibinfo{author}{L.~Jonke},
  \bibinfo{journal}{J. High Energy Phys.} \bibinfo{volume}{1112}
  (\bibinfo{year}{2011}) \bibinfo{pages}{080}.
  \DOIprefix\doi{10.1007/JHEP12(2011)080}.
\bibitem[{JuriÄ‡ et~al.(2013)JuriÄ‡, Meljanac, and
  Å trajn}]{EPJC.2013.73.2472}
\bibinfo{author}{T.~JuriÄ‡}, \bibinfo{author}{S.~Meljanac},
  \bibinfo{author}{R.~Å trajn}, \bibinfo{journal}{Eur. Phys. J. C}
  \bibinfo{volume}{73} (\bibinfo{year}{2013}) \bibinfo{pages}{2472}.
  \DOIprefix\doi{10.1140/epjc/s10052-013-2472-0}.
\bibitem[{Borowiec and Pachol(2009)}]{PRD.2009.79.045012}
\bibinfo{author}{A.~Borowiec}, \bibinfo{author}{A.~Pachol},
  \bibinfo{journal}{Phys. Rev. D} \bibinfo{volume}{79} (\bibinfo{year}{2009})
  \bibinfo{pages}{045012}. \DOIprefix\doi{10.1103/PhysRevD.79.045012}.
\bibitem[{Meljanac and StojiÄ‡(2006)}]{EPJC.2006.47.531}
\bibinfo{author}{S.~Meljanac}, \bibinfo{author}{M.~StojiÄ‡},
  \bibinfo{journal}{Eur. Phys. J. C} \bibinfo{volume}{47}
  (\bibinfo{year}{2006}) \bibinfo{pages}{531}.
  \DOIprefix\doi{10.1140/epjc/s2006-02584-8}.
\bibitem[{Meljanac et~al.(2008)Meljanac, Samsarov, StojiÄ‡, and
  Gupta}]{EPJC.2008.53.295}
\bibinfo{author}{S.~Meljanac}, \bibinfo{author}{A.~Samsarov},
  \bibinfo{author}{M.~StojiÄ‡}, \bibinfo{author}{K.~S. Gupta},
  \bibinfo{journal}{Eur. Phys. J. C} \bibinfo{volume}{53}
  (\bibinfo{year}{2008}) \bibinfo{pages}{295}.
  \DOIprefix\doi{10.1140/epjc/s10052-007-0450-0}.
\bibitem[{Andrade and Silva(2013)}]{PLB.2013.719.467}
\bibinfo{author}{F.~M. Andrade}, \bibinfo{author}{E.~O. Silva},
  \bibinfo{journal}{Phys. Lett. B} \bibinfo{volume}{719} (\bibinfo{year}{2013})
  \bibinfo{pages}{467}. \DOIprefix\doi{10.1016/j.physletb.2013.01.062}.
\bibitem[{Agostini et~al.(2004)Agostini, Amelino-Camelia, and
  Arzano}]{CQG.2004.21.2179}
\bibinfo{author}{A.~Agostini}, \bibinfo{author}{G.~Amelino-Camelia},
  \bibinfo{author}{M.~Arzano}, \bibinfo{journal}{Class. Quantum Grav.}
  \bibinfo{volume}{21} (\bibinfo{year}{2004}) \bibinfo{pages}{2179}.
  \DOIprefix\doi{10.1088/0264-9381/21/8/018}.
\bibitem[{Aloisio et~al.(2004)Aloisio, Galante, Grillo, Méndez, Carmona, and
  Cortés}]{JHEP.2004.2004.28}
\bibinfo{author}{R.~Aloisio}, \bibinfo{author}{A.~Galante},
  \bibinfo{author}{A.~F. Grillo}, \bibinfo{author}{F.~Méndez},
  \bibinfo{author}{J.~M. Carmona}, \bibinfo{author}{J.~L. Cortés},
  \bibinfo{journal}{J. High Energy Phys.} \bibinfo{volume}{2004}
  (\bibinfo{year}{2004}) \bibinfo{pages}{028}.
  \DOIprefix\doi{10.1088/1126-6708/2004/05/028}.
\bibitem[{Roy and Roychoudhury(1995)}]{PLB.1995.359.339}
\bibinfo{author}{P.~Roy}, \bibinfo{author}{R.~Roychoudhury},
  \bibinfo{journal}{Phys. Lett. B} \bibinfo{volume}{359} (\bibinfo{year}{1995})
  \bibinfo{pages}{339}. \DOIprefix\doi{10.1016/0370-2693(95)01079-6}.
\bibitem[{Roy and Roychoudhury(1994)}]{PLB.1994.339.87}
\bibinfo{author}{P.~Roy}, \bibinfo{author}{R.~Roychoudhury},
  \bibinfo{journal}{Phys. Lett. B} \bibinfo{volume}{339} (\bibinfo{year}{1994})
  \bibinfo{pages}{87}. \DOIprefix\doi{10.1016/0370-2693(94)91137-1}.
\bibitem[{Meljanac et~al.(2013)Meljanac, Pacho\l{}, Samsarov, and
  Gupta}]{PRD.2013.87.125009}
\bibinfo{author}{S.~Meljanac}, \bibinfo{author}{A.~Pacho\l{}},
  \bibinfo{author}{A.~Samsarov}, \bibinfo{author}{K.~S. Gupta},
  \bibinfo{journal}{Phys. Rev. D} \bibinfo{volume}{87} (\bibinfo{year}{2013})
  \bibinfo{pages}{125009}. \DOIprefix\doi{10.1103/PhysRevD.87.125009}.
\bibitem[{Gupta et~al.(2012)Gupta, Meljanac, and Samsarov}]{PRD.2012.85.045029}
\bibinfo{author}{K.~S. Gupta}, \bibinfo{author}{S.~Meljanac},
  \bibinfo{author}{A.~Samsarov}, \bibinfo{journal}{Phys. Rev. D}
  \bibinfo{volume}{85} (\bibinfo{year}{2012}) \bibinfo{pages}{045029}.
  \DOIprefix\doi{10.1103/PhysRevD.85.045029}.
\bibitem[{Govindarajan et~al.(2009)Govindarajan, Gupta, Harikumar, Meljanac,
  and Meljanac}]{PRD.2009.80.025014}
\bibinfo{author}{T.~R. Govindarajan}, \bibinfo{author}{K.~S. Gupta},
  \bibinfo{author}{E.~Harikumar}, \bibinfo{author}{S.~Meljanac},
  \bibinfo{author}{D.~Meljanac}, \bibinfo{journal}{Phys. Rev. D}
  \bibinfo{volume}{80} (\bibinfo{year}{2009}) \bibinfo{pages}{025014}.
  \DOIprefix\doi{10.1103/PhysRevD.80.025014}.
\bibitem[{Strange(1998)}]{Book.1998.Strange}
\bibinfo{author}{P.~Strange}, \bibinfo{title}{Relativistic Quantum Mechanics},
  \bibinfo{publisher}{Cambridge University Press}, \bibinfo{address}{Cambridge,
  England}, \bibinfo{year}{1998}.
\bibitem[{Abramowitz and Stegun(1972)}]{Book.1972.Abramowitz}
\bibinfo{editor}{M.~Abramowitz}, \bibinfo{editor}{I.~A. Stegun} (Eds.),
  \bibinfo{title}{Handbook of Mathematical Functions}, \bibinfo{publisher}{New
  York: Dover Publications}, \bibinfo{year}{1972}.

\end{thebibliography}

\end{document}